\documentstyle[psfig,epsf]{article}
\newcommand{\gcheck}{\mbox{$\check{g}$}}

\newcommand{\ghat}{\mbox{$\hat{g}$}}
\newcommand{\fhat}{\mbox{$\hat{f}$}}
\normalsize

\begin{document}
\title{Transport and triplet superconducting condensate in mesoscopic
ferromagnet-superconductor structures.}
\author{F. S. Bergeret $^{1}$, V.V.Pavlovskii$^{2}$, A. F. Volkov$^{1,2}$ and K. B.
Efetov$^{1,3}$\\$^{(1)}$Theoretische Physik III,\\
Ruhr-Universit\"{a}t Bochum, D-44780 Bochum, Germany\\
$^{(2)}$Institute of Radioengineering and Electronics of the Russian\\
Academy\\
of Sciences, 103907 Moscow, Russia \\
$^{(3)}$L.D. Landau Institute for Theoretical Physics, 117940 Moscow, Russia}

\maketitle
\begin{abstract}
We calculate the conductance  of a superconductor/ferromagnet (S/F)
mesoscopic structure in the dirty limit. First we assume that the ferromagnet exhibits a homogeneous magnetization and consider the case that the penetration of the condensate into the F wire is negligible  and the case in which the proximity effect is taken into account. 
It is shown that if the exchange field is large enough, the  conductance  below the critical temperature $T_C$, is always smaller than the conductance in the normal state. At last, we calculate the conductance for a F/S structure with a local inhomogeneity of the magnetization in the ferromagnet. We
demonstrate that a triplet component of the  condensate is induced in the F wire.This  leads to a  increase of the  conductance below $T_C$.
\end{abstract}

\section{Introduction}

In the last decade transport properties of mesoscopic superconductor/normal
metal (S/N) structures were intensively studied (see for example the review
articles \cite{beenakker_rev,lambert_rev} and references therein). It was
established that in these nano-structures, i.e. in structures whose
dimensions are less than the phase coherence length $L_{\varphi }$ and the
inelastic scattering length $l_{\varepsilon }$, the conductance changes when
decreasing the temperature below the critical temperature $T_{c}$ of the
superconducting transition and this variation may be both positive ($\delta
G>0$) and negative ($\delta G<0$) \cite
{petrashov_91,vanWees_95,charlat,chien1,chien2,shapira,wilhelm}. The increase or
decrease of the conductance $G$ depends, in particular, on the interface
resistances and is determined by a competition between two contributions
caused by the proximity effect. One of them is due to the suppression of the
density-of-state (DOS) and leads to a decrease of the conductance. The
second one results in increasing $G$ and is similar to the Maki-Thompson
term \cite{golubov_zaikin,volkov_pavlo}.

These studies apparently were stimulated by the theoretical work \cite
{spivak_khme} in which a weak-localization correction to the conductance of
a S/N/S mesoscopic structure was calculated and an oscillatory dependence of
this correction on the phase difference $\varphi $ between the
superconductors S was predicted. These oscillations have been indeed
observed \cite{vanWees_95,charlat,chien2,petrashov_sn,pothier,vegvar,nitta}
but their amplitude turned out to be two orders of magnitude larger than the
predicted one. In order to overcome this discrepancy between theory and
experiment one should take into account the proximity effect. It was shown
that the latter leads to much larger amplitudes of the conductance
oscillations than the weak-localization corrections \cite
{hekking_93,zaitsev_a}.

The proximity effect manifests itself also in other interesting
peculiarities of the transport properties of S/N structures. One of them is
an interesting dependence of the Josephson current $I_{c}$ on an additional
dissipative current $I_{ad}$ through a N wire in a 4-terminal S/N/S
structure similar to the one shown in the inset of Fig.\ref{fig_geometry_triplet}. According to theoretical predictions the current $%
I_{c}$ changes sign if the current $I_{ad}$ is large enough \cite
{vanWees_91,volkov,yip,wilhelm_schon,wendin}. This behavior of the Josephson
current ($\pi $-contact) was later experimentally confirmed\cite{baselmans}.
Another interesting effect is a non-monotonic temperature $T$ (or voltage $V$%
) dependence of the correction to the conductance $\delta G$ \cite
{gubankov,charlat,chien1,chien2,shapira}. When decreasing the temperature, $%
\delta G$ increases, reaches a maximum and with further decrease of the
temperature drops to zero. This behavior has been explained theoretically
for the case of a short S/N contact ($E_{Th}>>\Delta $) \cite{artemenko}and
a ''long'' S/N structure ($E_{Th}<<\Delta $) \cite
{nazarov_stoof,volkov_allsopp}, here $E_{Th}=\hbar D/L^{2}$ is the Thouless
energy, $D$ is the diffusion constant, $L$ is the length of the N wire
(film). The reason for this behavior is the competition between two
contributions to the conductance mentioned above.

It is interesting to note that low-energy states play an important role in
transport properties of S/N structures. The reason is that the condensate
penetrates into the N wire over a length $\xi _{\varepsilon }=\sqrt{\hbar
D/\varepsilon }$ which may be much larger than the thermodynamic correlation
length $\xi _{T}=\sqrt{\hbar D/2\pi T}$ provided the characteristic energy $%
\varepsilon \cong E_{Th}<<T.$ Therefore, in the limit of low (compared to $T$%
) Thouless energy $E_{Th},$ the phase coherence is maintained over distances 
$\xi _{\varepsilon }=\sqrt{\hbar D/E_{Th}}$ of the order of the length $L$
of the N wire. These long-range phase-coherent effects have been predicted
in \cite{volkov_pavlo_jetp,zhou} and observed on a 4-terminal S/N/S structure 
\cite{courtois,kutchinsky}. In Ref.\cite{wilhelm} the conductance oscillations related to the phase coherence 
were observed in a S/N/S mesoscopic structure.
It was established for the first time that these oscillations survive in 
a temperature range where the Josephson
coupling between the superconductors becomes negligible. The authors of 
Ref.\cite{kutchinsky} observed an increase in the Josephson
critical current when an additional dissipative current was injected 
into the N region in a 4-terminal S/N/S structure.
This current leads to long-range effects affecting the critical current.

Recently, similar investigations have been carried out on mesoscopic F/S
structures in which ferromagnets (F) are used instead of normal
(nonmagnetic) metals. It is well known that in the dirty limit, when the
relaxation momentum time $\tau $ is very small ($\tau <<\hbar /h_{ex}$,
where $h_{ex}$ is the exchange energy), the condensate penetrates into a
ferromagnet over a length $\xi _{F}=\sqrt{\hbar D/h_{ex}}$. The latter is
extremely short (5-50 \AA ) for\ strong ferromagnets like Fe or Ni.
Therefore one might expect that the influence of the proximity effect on
transport properties of such structures should be negligibly small.
Experiments carried out recently on F/S structures showed however that the
conductance variation $\delta G$ is quite visible (varying from about 1 to
10\%) when the temperature decreases below $T_{c}$ \cite
{petrashov,chandrasekhar,pannetier}. It is worth mentioning that the
conduction variation was both positive and negative. In some experiments the
variation $\delta G$ was related to a variation of the interface conductance
(resistance) \cite{chandrasekhar}, whereas in others \cite{petrashov} to the
conduction variation of the ferromagnetic wire $\delta G_{F}$. The theory
also predicts both an increase \cite{dejong,golubov,nazarov_belzig} and
decrease\cite{vanWees_99,falko_volkov1} of the conductance.

In Refs. \cite{dejong,golubov,nazarov_belzig} a ballistic contact was
analyzed. It was shown that at $h_{ex}=0$ the contact conductance $G_{F/S}$
is twice as large as its conductance $G_{F/N}$ in the normal state (above $%
T_{c}$), which agrees with a theoretical prediction for a N/S ballistic
contact, and drops to zero at $h_{ex}=E_{F},$ where $E_{F}$ is the Fermi
energy. The conductance of a diffusive point contact $G_{F/S}$ has been
calculated by Golubov \cite{golubov} who showed that $G_{F/S}$ is always
smaller than the conductance $G_{F/N}$ in the normal state. In the case of a
mixed conductivity mechanism (partly diffusive and partly ballistic) $%
G_{F/S} $ has been calculated in Ref. \cite{nazarov_belzig} and it may be
both larger or smaller than the conductance in the normal state $G_{F/N}$.

In order to obtain the resistance of the system shown in Fig.\ref
{fig_geometry_triplet} one should add the interface resistances $R_{b}$ and
the resistance of the F wire $R_{F}$. The conductance of the F wire $%
G_{F}=R_{F}^{-1}$ under the assumption that $R_{b}$ is sufficiently
small has been calculated in Refs. \cite{vanWees_99,falko_volkov1}. As in
Ref.\cite{nazarov_belzig}, the proximity effect was neglected. It turned out
that the conductance of the F wire in the normal state was $%
G_{Fn}=G_{\uparrow }+G_{\downarrow }$ and was equal to $G_{Fs}=4G_{\uparrow
}G_{\downarrow }/(G_{\uparrow }+G_{\downarrow })$ in the superconducting
state (these formulas are valid for a F wire shorter than the spin
relaxation length, see \cite{falko_volkov1}). Thus, the conductance of the F
wire decreases below $T_{c}$. The mechanism responsible for this behavior is
in this case purely kinetic, since the form of the distribution function
depends only on the boundary conditions at the interfaces: if the interface
transparencies are perfect the distribution function is continuous in the
normal state, while in the superconducting state the boundary condition is
changed providing the spinless current through the F/S interface. These two
different boundary conditions for the distribution function lead to the
different values of the conductance above and below $T_{c}$.
In the present paper we analyze the conductivity of a F/S structure in the
dirty limit when the condition $\tau <<\hbar /h_{ex}$ is satisfied.

In the next section we study the kinetic mechanism of the conductance
variation in a F/S mesoscopic structure neglecting the proximity effect. The
diffusion coefficients and the interface transparencies are assumed to be
different for each spin direction. It will be shown that in this case the
conductance variation is always negative, {\it i.e.} the conductance
decreases with decreasing temperature below $T_{c}$.

In the third section we take into account the proximity effect (the
condensate penetration into the ferromagnetic wire) and present the results
for the conductance on the basis of exact calculations assuming equal
diffusion constants and interface transparencies for spin up and down. We
will see that for a strong exchange field $E_{exc}>>T_{c}$, the resistance
variation $\delta R$ is positive at any temperature and interface resistance.

Finally, in section 4 we calculate the conductance variation assuming that
the magnetization near the F/S interface has a local inhomogeneity. It will
be shown that not only a singlet component arises in the F wire in this case
but also a triplet one which penetrates into the F wire over a long distance
of the order of $\xi _{m}=\sqrt{D/\epsilon _{m}}$, where $\epsilon _{m}=\min
\{T,E_{Th}\}$. The penetration of the triplet component leads to a positive $%
\delta G$ which may be comparable with the conductance variation observed in
N/S structures. The singlet component exists only in a very short region
near the F/S interface and leads to a negligible contribution to the
conductance.

\section{ Kinetic mechanism of the conductance variation}

%%%%%%%%%%%%%%%%%
%\begin{figure}
%\epsfysize= 3.5cm
%\vspace{0.2cm}
%\centerline{\epsfbox{./figures/cond_geometry1.eps}}
%\vspace{0.2cm}
%\caption{S/F system.}
%\label{fig_geometry1}
%\end{figure}
%%%%%%%%%%%%%%%%%

\begin{figure}
\epsfysize= 6cm
\vspace{0.2cm}
\centerline{\epsfbox{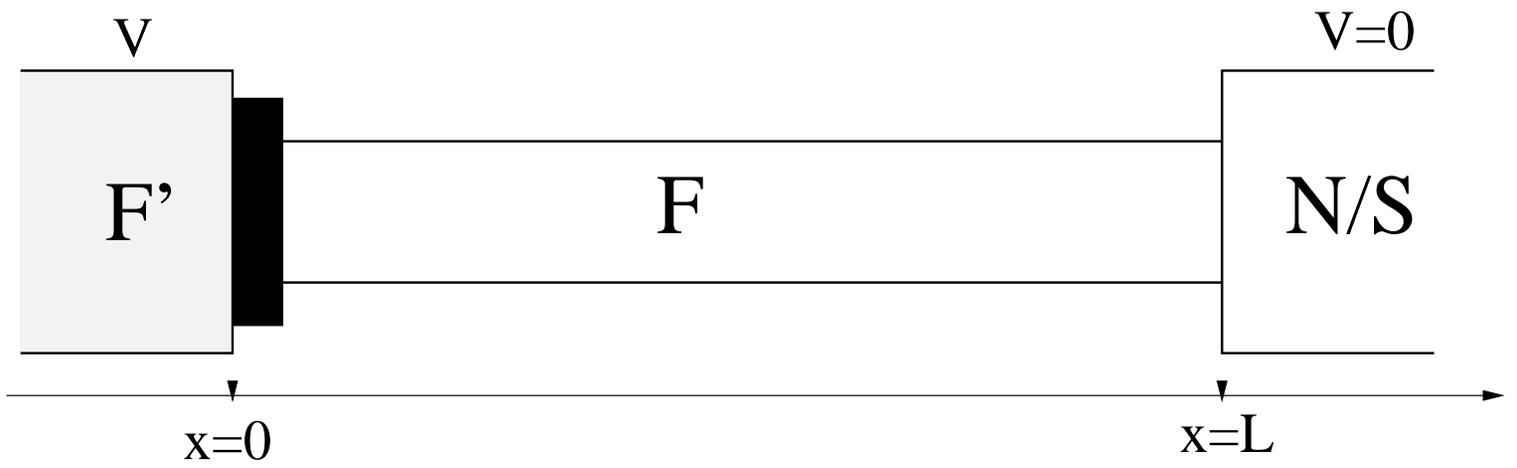}}
\vspace{2cm}
\caption{F. S. Bergeret, A.F. Volkov and K.B.Efetov. ``Transport and triplet superconducting condensate in mesoscopic
ferromagnet-superconductor structures.''}
\label{fig_geometry1}
\end{figure}

In this section, we calculate the conductance of the system shown in Fig.\ref
{fig_geometry1}. We neglect influence of the proximity effect on the
conductance assuming that the penetration length of the condensate is very
small in comparison with the length of the F wire. The penetration length of
the condensate is equal to $\xi _{F}=\sqrt{\hbar D/h_{ex}}$ in the dirty
limit and to the mean free path $l=v_{F}\tau $ in the ''clean'' limit (to be
more exact, in the limit of a strong ferromagnet, when the condition $\tau
>>\hbar /h_{ex}$ is satisfied) . The conduction variation is related in this
case to different forms of the distribution function in the normal and
superconducting states. The form of the distribution function is determined
by boundary conditions at the interfaces F/N ($T>T_{c}$) or F/S ($T<T_{c}$).
In the normal state the spin current is finite, whereas in the
superconducting one it is equal to zero. In the superconducting state at low
temperatures, charge is transferred via Andreev reflections, i.e. an
incident electron is reflected from the F/S boundary and moving back as a
hole with the opposite spin direction (in the S region the charge is carried
by singlet Cooper pairs with zero total spin). In order to find the charge
or spin current we need to calculate the distribution function $N_{\alpha
}=n_{\alpha }(\epsilon )-p_{\alpha }(\epsilon )$ obeying the diffusion
equation

\begin{equation}  \label{eq_1}
D_{\alpha } \partial_{xx}^{2}N_{\alpha }=I_{in}(N_{\alpha }).
\end{equation}

Here $n_{\alpha }(\epsilon )$ is the distribution function of electrons for
a given spin direction $\alpha $, $p_{\alpha }(\epsilon )=1-n_{\alpha
}(-\epsilon )$ is the distribution function of holes, $D_{\alpha }$ is the
diffusion coefficient which depends on the spin index $\alpha ,$ $%
I_{in}(N_{\alpha })$ is the inelastic collision integral. The inelastic
collision integral is of the order of $N_{\alpha }/\tau _{in}$, where $\tau
_{in}$ is the inelastic scattering time. We ignore spin relaxation processes
in the F wire assuming that the spin relaxation length exceeds the length $L$%
. It will be seen that the conductance in the superconducting state is
always larger than in the normal one. The simplest formulas are obtained in
the case of a perfect F/S interface. Therefore, we assume that there is no
barrier at the F/S interface and the F'/F interface transparency is
arbitrary (F' is the ferromagnetic or normal reservoir). In the next section
we will show that the non-zero reflection coefficient at the F/S interface
leads also to an increase of the resistance when temperature is lowered
below $T_{c}.$

The boundary conditions at the F'/F interface have the same form in the
normal and superconducting states (see Appendix)

\bigskip 
\begin{equation}
D_{\alpha } \partial _{x}N_{\alpha }=(\gamma R_{b})_{_{\alpha
}}^{-1}(N_{\alpha }(0)-F_{V});\ \ x=0  \label{K1}
\end{equation}

Here $N_{\alpha }(0)=N_{\alpha }(x=0),$ $R_{b\alpha }$ is the barrier
resistance per unit area for a given spin direction $\alpha $, $\gamma
=e^{2}\nu ,$ $\nu $ is the density-of-states (DOS). The function $%
F_{V}=[\tanh ((\epsilon +eV)/2T)-\tanh ((\epsilon -eV)/2T)]/2$ is the
distribution function in the F' reservoir. At the F/S interface the boundary
conditions are different in the normal and superconducting case. Above $%
T_{c} $ they are

\begin{equation}
N_{\alpha }=0;\ \ x=L  \label{K2}
\end{equation}

Below $T_{c}$ they have the form (see Appendix)

\begin{equation}
N_{\alpha }=-N_{\bar{\alpha}};\ \ x=L  \label{K3}
\end{equation}

\begin{equation}
\nu_{\alpha }D_{\alpha } \partial _{x}N_{\alpha }= \nu_{\bar{\alpha}}D_{\bar{\alpha}}  \partial_{x}N_{\bar{%
\alpha}}; \;\; x=L,\;\; |\epsilon |<\Delta  \label{K4}
\end{equation}

Physically, Eqs.(\ref{K2}) and (\ref{K3}) mean that the electric potential
at the perfect F/N(S) interface is chosen to be zero. The potential $V(x)$
is expressed in terms of $N_{\alpha }$ as \cite{larkin_ovch_book} $%
eV(x)=(1/4)\int {\rm d}\epsilon \lbrack N_{\uparrow }+N_{\downarrow }]$. The
condition (\ref{K3}) implies that the spin current is zero. This fact is a
result of the Andreev reflections. It is worth mentioning that, in the model
under consideration, the conductance related to the Andreev reflections
differs from zero only if the amplitude of the condensate function at the
F/S interface is not equal to zero. Therefore, strictly speaking, we neglect
the proximity effect everywhere in the F wire except the nearest
neighborhood of the F/S interface. The electrical current is given by the
formula

\begin{equation}
I_{Q}=(1/4e)\int {\rm d}\epsilon \lbrack (\gamma D)_{\alpha } \partial
_{x}N_{\alpha }+(\gamma D)_{\bar{\alpha}} \partial _{x}N_{\bar{\alpha}}]
\label{K5}
\end{equation}

The spin current $I_{M}$ is given by the same formula with sign ``-''
instead of ``+'' in the brackets and a factor of $g\mu/e $ in front of the
integral ($\mu $ is the Bohr magneton). The problem is reduced to finding a
solution for Eq.(\ref{K1}) with the boundary conditions (\ref{K2}-\ref{K4}).
The inelastic collision integral has a complicated form and finding a
solution is in  general  an unrealistic task. Therefore, we consider
limiting cases only.

\subsection{ Mesoscopic limit ($L<<l_{in}$)}

In this case of a long inelastic relaxation length $l_{in}=\sqrt{D\tau _{in}}
$, we can neglect the inelastic collision integral and seek a solution in
the form

\begin{equation}
N_{\alpha }(x)=N_{\alpha }(0)-J_{\alpha }x/L\;.  \label{K6}
\end{equation}
where $J_{\alpha }=N_{\alpha }(L)-N_{\alpha }(0)$ is a constant that
determines the current and, hence, the conductance. The constants $N_{\alpha
}(0)$ and $J_{_{\alpha }}$ can be determined from the boundary conditions (%
\ref{K2}-\ref{K4}) and are equal to

\begin{equation}
J_{\alpha }=F_{V}\left( \frac{G_{b}}{G_{L}+G_{b}}\right) _{\alpha
},\;\;N_{\alpha }(0)=J_{\alpha }L\;.  \label{K7}
\end{equation}
in the normal case, and

\begin{equation}
J_{\alpha }=F_{V}\frac{2\mbox{$G_{L\bar{\alpha}}$}}{\mbox{$G_{L\alpha}$}%
\mbox{$G_{L\bar{\alpha}}$}(\mbox{$R_{b\alpha}$}+\mbox{$R_{b\bar{\alpha}}$})+%
\mbox{$G_{L\alpha}$}+\mbox{$G_{L\bar{\alpha}}$}},\;\;N_{\alpha }(0)=F_{V}%
\frac{\mbox{$G_{L\alpha}$}\mbox{$G_{L\bar{\alpha}}$}(\mbox{$R_{b\bar{%
\alpha}}$}-\mbox{$R_{b\alpha}$})}{\mbox{$G_{L\alpha}$}\mbox{$G_{L\bar{%
\alpha}}$}(\mbox{$R_{b\alpha}$}+\mbox{$R_{b\bar{\alpha}}$})+(%
\mbox{$G_{L\alpha}$}+\mbox{$G_{L\bar{\alpha}}$})}\;.  \label{K8}
\end{equation}
in the superconducting case, where $R_{L\alpha}=\rho_{\alpha}L$ is the resistance of the F wire in the normal state. Knowing the distribution function we can
compute the conductances and the spin current $I_{M}$. In the normal state,
we obtain

\begin{equation}
G_{n}=\left( \frac{G_{b}G_{L}}{G_{b}+G_{L}}\right) _{\uparrow }+\left( \frac{%
G_{b}G_{L}}{G_{b}+G_{L}}\right) _{\downarrow },\;\;I_{M}=(g\mu/e)V\left[ \left( 
\frac{G_{b}G_{L}}{G_{b}+G_{L}}\right) _{\uparrow }-\left( \frac{G_{b}G_{L}}{%
G_{b}+G_{L}}\right) _{\downarrow }\right] \;.  \label{K9}
\end{equation}

whereas the result for the superconducting state reads

\begin{equation}
G_{s}=\frac{4\mbox{$G_{L\uparrow}$}\mbox{$G_{L\downarrow}$}%
\mbox{$G_{b\uparrow}$}\mbox{$G_{L\downarrow}$}}{\mbox{$G_{L\uparrow}$}%
\mbox{$G_{L\downarrow}$}(\mbox{$G_{b\uparrow}$}+\mbox{$G_{b\downarrow}$})+(%
\mbox{$G_{L\uparrow}$}+\mbox{$G_{L\downarrow}$})\mbox{$G_{b\uparrow}$}%
\mbox{$G_{b\downarrow}$}},\;\;I_{M}=0\;.  \label{K10}
\end{equation}

As follows from Eqs.(\ref{K9}-\ref{K10}), the spin current in the normal
state is absent only if the conductances in each spin band $G_{b\alpha }$
and $G_{L\alpha }$ \ are the same. In the superconducting state this current
is always zero. The conductances $G_{b\alpha }$ and $G_{L\alpha }$ enter the
formulas for $G_{n}$ and $G_{s}$ (\ref{K9}-\ref{K10}) symmetrically. In the
case of a small F'/F interface resistance, we obtain for $G_{n}$ and $G_{s}$
the formulas presented in \cite{falko_volkov1}and in the introduction. The
conductance in the normal state $G_{n}$ is always larger than or equal to
(if conductances in each spin channel are the same) the conductance in
superconducting state $G_{s}$. Indeed, the difference between $G_{n}$ and $%
G_{s}$ can be written in the form

\begin{equation}
G_{n}-G_{s}=\left[ \mbox{$G_{L\uparrow}$}\mbox{$G_{L\downarrow}$}(%
\mbox{$G_{b\uparrow}$}-\mbox{$G_{b\downarrow}$})+\mbox{$G_{b\uparrow}$}%
\mbox{$G_{b\downarrow}$}(\mbox{$G_{L\uparrow}$}-\mbox{$G_{L\downarrow}$})%
\right] ^{2}/{\cal D}_{F/N}{\cal D}_{F/S},\;.  \label{K12}
\end{equation}
where ${\cal D}_{n}$ and ${\cal D}_{s}$ are the denominators in the
expressions (\ref{K9}-\ref{K10}). Therefore, in the model under
consideration (dirty limit, no condensate penetration, zero F/S interface
resistance) the conductance decreases with decreasing temperature below $%
T_{c}$.

Note that the formula for the conductance Eq.(\ref{K10}) is valid at low
temperatures ($T<<\Delta $) when the contribution of the states with 
%TCIMACRO{\TEXTsymbol{\vert}}%
%BeginExpansion
\mbox{$\vert$}%
%EndExpansion
$\epsilon $%
%TCIMACRO{\TEXTsymbol{\vert}}%
%BeginExpansion
\mbox{$\vert$}%
%EndExpansion
%TCIMACRO{\TEXTsymbol{>}}%
%BeginExpansion
\mbox{$>$}%
%EndExpansion
$\Delta $ to the conductance can be neglected. At arbitrary temperatures the
conductance can be easily found with the help of Eq.(\ref{K5}) and has the
form

\begin{equation}
G_{s}(T)=G_{s}(0)\tanh (\Delta /2T)+G_{n}(1-\tanh (\Delta /2T))
\end{equation}
where $G_{s}(0)$ is the conductance of the F/S structure at zero temperature
determined by Eq.(\ref{K10}). We took into account that at \ 
%TCIMACRO{\TEXTsymbol{\vert}}%
%BeginExpansion
\mbox{$\vert$}%
%EndExpansion
$\epsilon |>\Delta$ the boundary conditions for the distribution function
are given by Eq.(3).

\subsection{ Semi-mesoscopic limit ($L>>l_{in}$)}

Let us consider another limiting case when the length $L$ of the F-wire
exceeds the energy relaxation length $l_{in}.$ In this case the inelastic
scattering term in Eq.(\ref{eq_1}) dominates. It turns to zero provided the
distribution function has the equilibrium form

\begin{equation}
N_{\alpha }(x)=[\tanh ((\epsilon +eV_{\alpha }(x))/2T)-\tanh ((\epsilon
-eV_{\alpha }(x))/2T)]/2\;.  \label{K13}
\end{equation}
where the potential $V_{\alpha }$ depends on the spin index $\alpha $ and is
linear as a function of the coordinate $x$ : $V_{\alpha }(x)=V_{\alpha
}(0)-E_{\alpha }x$. Integrating Eq. (\ref{K1}) over the energies we obtain a
relation between the current in the $\alpha $ spin band and $V_{\alpha }(0)$

\begin{equation}
I_{\alpha }=G_{b\alpha }(V-V_{\alpha }(0))\;.  \label{K14}
\end{equation}
Let us consider first the normal case. In the F wire we have

\begin{equation}
I_{\alpha }=G_{L\alpha }V_{\alpha }(0)/L\;.  \label{K15}
\end{equation}
From the last two equations we find $V_{\alpha }(0)$

\begin{equation}
V_{\alpha }(0)=V\frac{G_{b\alpha }}{G_{b\alpha }+G_{L\alpha }}\;.
\label{K16}
\end{equation}

The electric field and the total current are equal to the sum of $E_{\alpha
} $ and $I_{\alpha }$ over spin indices respectively. We easily obtain an expression for the conductance $G_{n}$, which is identical to Eq. (\ref
{K9}). In the superconducting case the current $I_{\alpha }$ is given by

\begin{equation}
I_{\alpha }=G_{L\alpha }(V_{\alpha }(0)-V_{\alpha }(L))/L\;.  \label{K17}
\end{equation}

The potentials $V_{\uparrow }(L)$ and $V_{\downarrow }(L)$ are related to
each other by Eq. (\ref{K3}): $V_{\uparrow }(L)=-V_{\downarrow }(L)$.
Although the form of the distribution function differs from Eqs.(\ref{K6}-%
\ref{K7}), the conductance $G_{s}$ below $T_{c}$ remains unchanged  with
respect to the previous case (see Eq. (\ref{K10})). Thus, regardless of the
relationship between the lengths $L$ and $l_{in}$, the conductance of the
structure under consideration is given by Eqs. (\ref{K9}-\ref{K10}), {\it i.e%
} it decreases with decreasing temperature below $T_{c}$.

\section{Singlet component and proximity effect}

In the preceding section we neglected the proximity effect and assumed that
the F/S interface resistance was equal to zero. We have shown that the
conductance decreases with decreasing temperature provided the
conductivities of the ferromagnetic wire or the F'/F interface for spin up
and down differ from each other. In the present section we calculate the
conductance for an arbitrary F/S interface resistance taking into account
the proximity effect (the penetration of the superconducting condensate into
the F wire). In this case the problem becomes more complicated and, for
simplicity, we assume that conductivities for the both spin directions are
the same. An equation for the distribution function can be solved exactly
and this allows one to present the normalized zero-bias conductance $%
S=R_{F/N}(\partial I/\partial V)$ in the form

\begin{equation}
S=(1/2)\int d\epsilon \frac{(1+r_{L}+r_{0})\partial F_{V}/\partial V}{%
\langle 1/M\rangle +r_{0}/\nu (0)+r_{L}/(g_{QP}+g_{A})}.  \label{K18}
\end{equation}
where $r_{L,0}=R_{b}(L,0)/R_{L}$ are the normalized resistances of the F/S
and F'/F interfaces respectively (in the normal state), $%
2M(x)=1+|f^{R}(x)|^{2}+|g^{R}(x)|^{2},\nu (0)=%
%TCIMACRO{\func{Re}}%
%BeginExpansion
\mathop{\rm Re}%
%EndExpansion
g^{R}(0),$ $\nu _{S}=%
%TCIMACRO{\func{Re}}%
%BeginExpansion
\mathop{\rm Re}%
%EndExpansion
g_{S}^{R},g_{S}^{R}=\epsilon /\sqrt{\epsilon ^{2}-\Delta ^{2}}$ is the
retarded Green's function in the superconductor, $g_{QP}=\nu (L)\nu _{S}$ is
the quasiparticle normalized conductance at a given energy, $g_{A}=%
%TCIMACRO{\func{Im}}%
%BeginExpansion
\mathop{\rm Im}%
%EndExpansion
f^{R}(L)%
%TCIMACRO{\func{Im}}%
%BeginExpansion
\mathop{\rm Im}%
%EndExpansion
f_{S}^{R}$ is the normalized subgap conductance (the conductance related to
Andreev reflections), $f_{S}^{R}=\Delta /\sqrt{\epsilon ^{2}-\Delta ^{2}}$.
The angle brackets mean the averaging over the length: $\langle ...\rangle
=\int_{0}^{1}dx/L(...).$ The physical meaning of the denominator is simple:
the first term in the angle brackets is the normalized resistance of the F
wire in the presence of the condensate, the second and third terms are the
resistances of the F'/F and F/S interfaces below $T_{c}$ respectively. One
can see that the ''Andreev'' conductance $g_{A}$ is not zero only if the
condensate function at the interface (from the ferromagnetic side) $F^{R}(L)$
differs from zero. The factor $(1+r_{L}+r_{0})$ in the numerator arises from
the normalization of the conductance. In order to calculate the normalized
conductance $S$, we have to find the Green functions $f^{R}(x)$ and $g^{R}(x)
$ in the ferromagnetic wire. These functions obey the Usadel equation which
upon the substitution $g^{R}(x)=\cosh u(x)$ and $f^{R}(x)=\sinh u(x)$
acquires the well known form

\begin{equation}
D\partial _{xx}^{2}u+2i(h_{ex}+\epsilon )\sinh u=0\;.  \label{K19}
\end{equation}

This equation is complemented by the boundary conditions

\begin{equation}
r_{0}L\partial _{{\bf r}}u=\sinh u,\;\; x=0  \label{K20}
\end{equation}

\begin{equation}
r_{L}L\partial _{{\bf r}}u=F_{S}^{R}\cosh u-G_{S}^{R}\sinh u,\;\; x=L  \label{K21}
\end{equation}

\begin{figure}
\epsfysize = 14cm
\vspace{0.2cm}
\centerline{\epsfbox{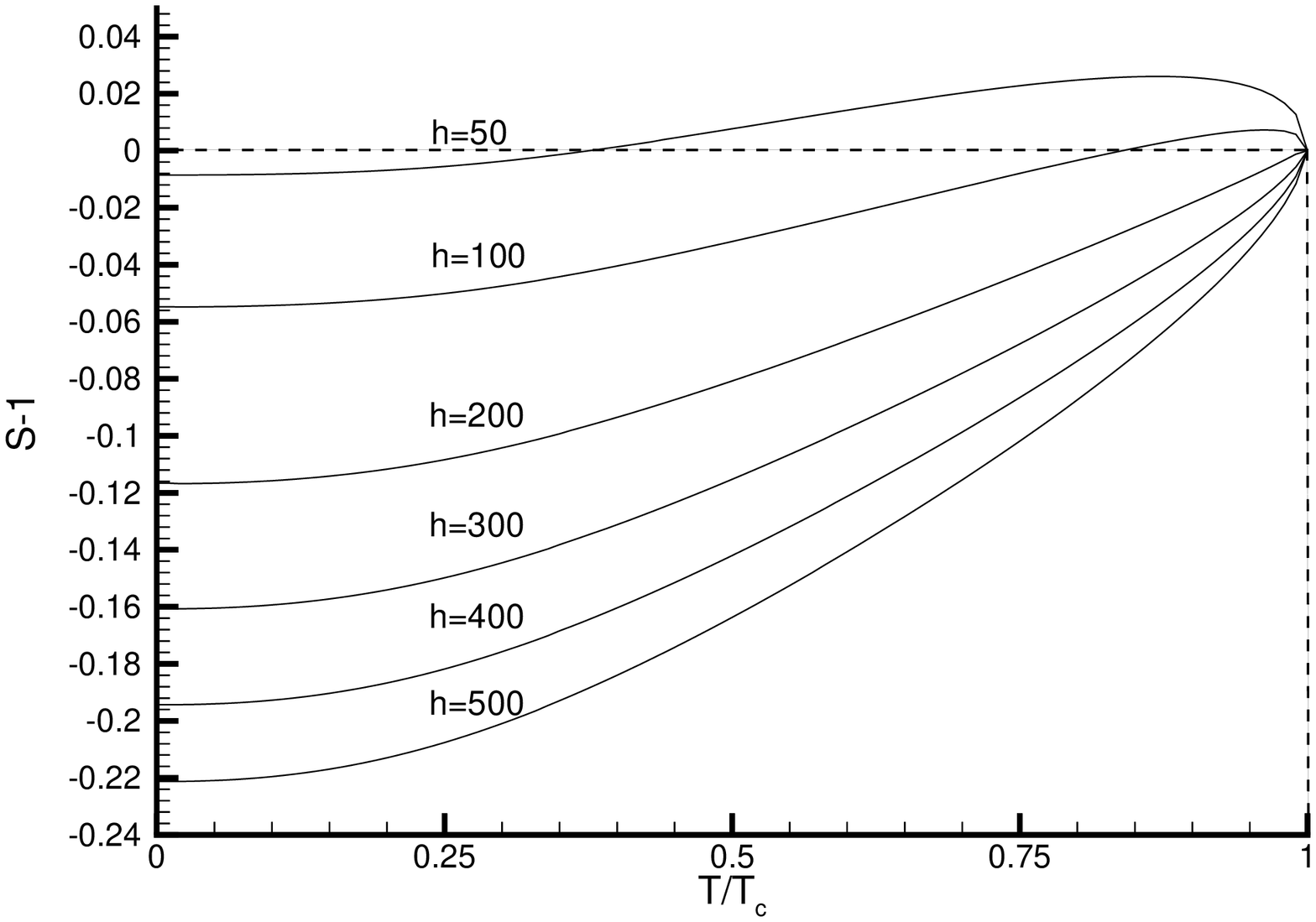
 }}
\vspace{2cm}
\caption{F. S. Bergeret, A.F. Volkov and K.B.Efetov. ``Transport and triplet superconducting condensate in mesoscopic
ferromagnet-superconductor structures.''}
\label{fig_plot1}
\end{figure}

The solution for Eq.(\ref{K19}) can be found analytically in some limiting
cases \cite{golubov_zaikin,VZK}. It has an especially simple form in the
case of a weak proximity effect when $r_{L}\leq 1$ and Eq.(19) may be
linearized (numerical calculations show that the linearized solution differs
from the exact solution by less than 10\% for $r_{L}\cong 1$). Here we
present the results of a numerical solution of the Usadel equation Eq.(\ref
{K19}) and of the calculation of the conductance $S.$ In Fig.\ref{fig_plot1}
the temperature dependence of the conductance variation $\delta S=S-1$ is
presented for a structure with good interface transparencies at various
exchange energies $h$ (normalized to the Thouless energy $\hbar D/L^{2}$). 

It is seen from this figure that $\delta S$ is positive only if $h_{ex}$ is
not too large compared to $\Delta $ or $T_{c}$. A small positive value $%
\delta S$ is observed near $T_{c}$ for $h_{ex}/\Delta =5$ and $10.$ For $%
h_{ex}/\Delta \leq 20$ the conductance variation is negative and
decreases with decreasing $T.$ For example, in the case of F/Al structure
the threshold exchange energy $h=20\Delta $ above which the conductance
deviation is negative for all temperatures, is equal to $h_{ex}=44$ $K$,
that is, this value is much less than the characteristic magnitudes of $%
h_{ex}$ for ferromagnets as Fe, Ni,Co etc. 

\begin{figure}[tbp]
\epsfysize = 14cm
\vspace{0.2cm}
\centerline{\epsfbox{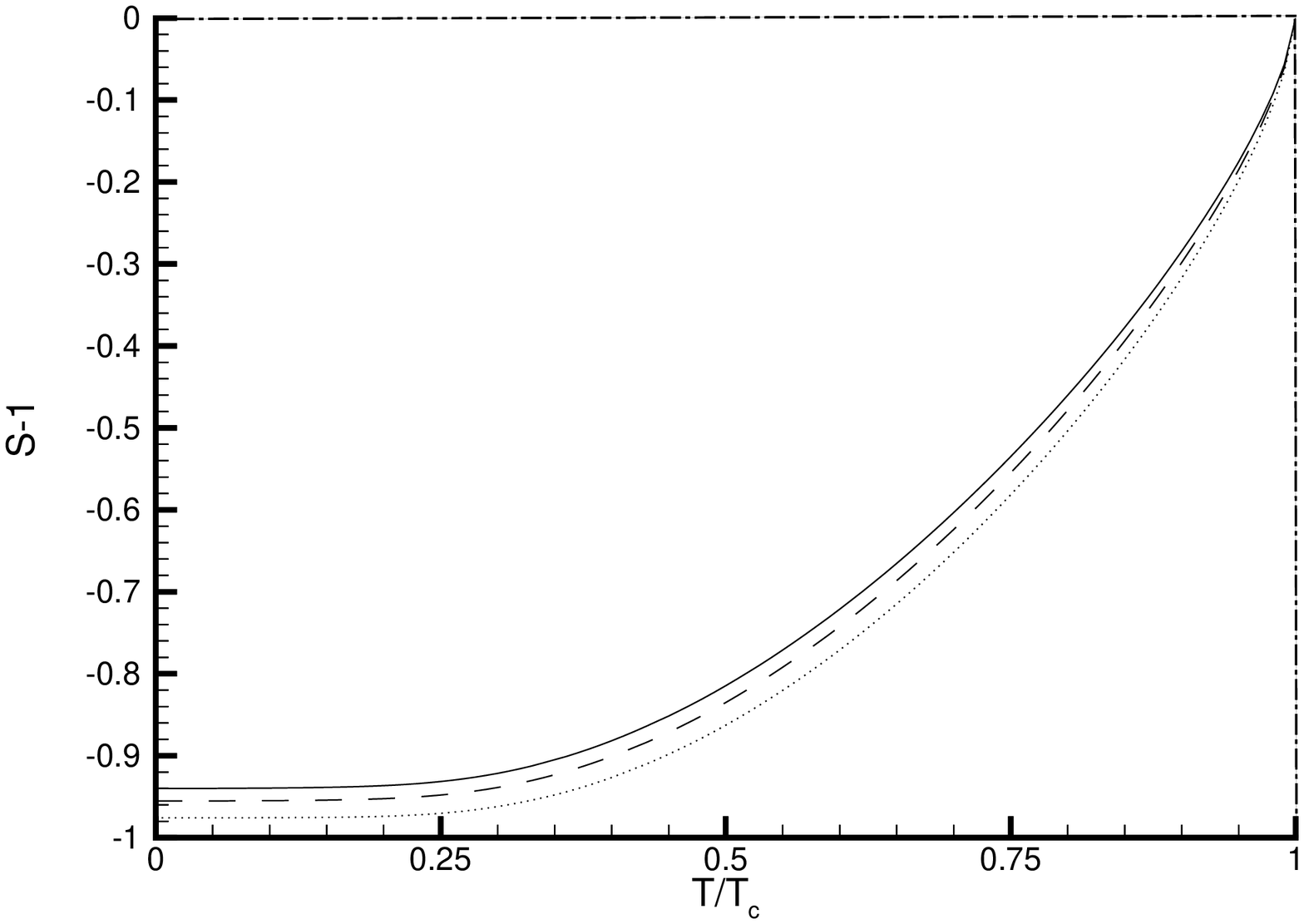
 }}
\vspace{2cm}\caption{F. S. Bergeret, A.F. Volkov and K.B.Efetov. ``Transport and triplet superconducting condensate in mesoscopic
ferromagnet-superconductor structures.''}
\label{fig_plot2}
\end{figure}

\begin{figure}[tbp]
\epsfysize = 10cm
\centerline{\epsfbox{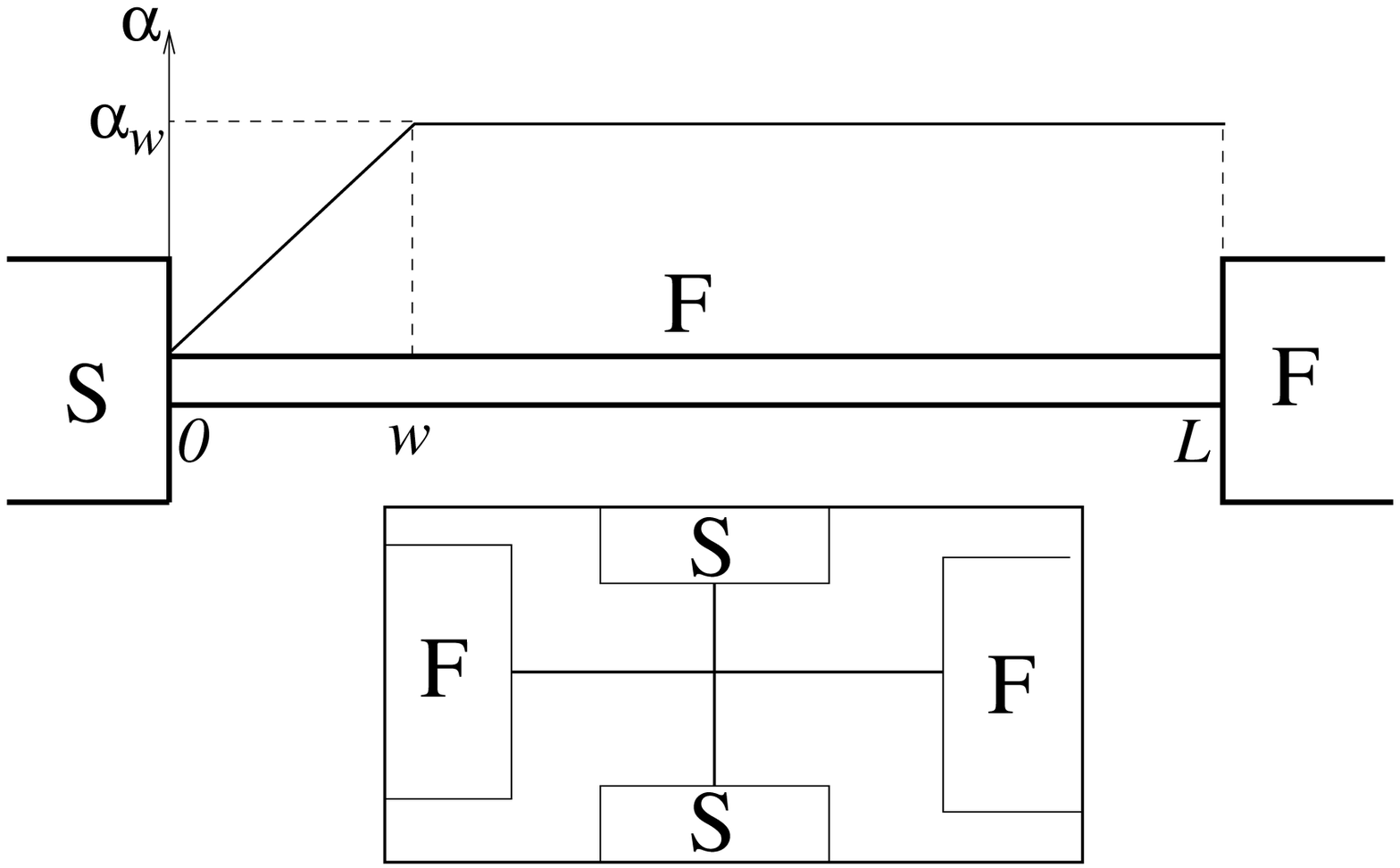
 }}
\vspace{2cm}
\caption{F. S. Bergeret, A.F. Volkov and K.B.Efetov. ``Transport and triplet superconducting condensate in mesoscopic
ferromagnet-superconductor structures.''}
\label{fig_geometry_triplet}
\end{figure}

A slight increase of $G_{s}$ above $G_{n}$ is related to the condensate
penetration into the F wire. This increase occurs only near $T_{c}$ because
in this temperature range the conductance $G_{s}$ is close to the
conductance $G_{F/N}$ in the normal state due to a large contribution of the
quasiparticle current. When the temperature decreases, the quasiparticle
contribution also decreases and the weak (at large $h$) proximity effect can
not overcome this decrease of the conductance. The contribution to the
conductance due to proximity effect is suppressed even stronger if the F/S
interface resistance is not small compared to $R_{L}$. In Fig. \ref
{fig_plot2} we plot the temperature dependence of $\delta S$ for the case
when the F/S interface resistance (in the normal state) is two times larger
than the conductance of the F wire. One can see that the normalized
conductance decreases with decreasing temperature and becomes very small at
low $T.$ The reason for this behavior of the conductance is quite clear. At
high enough temperatures the main contribution to the conductivity is due to
quasiparticles with energies $\epsilon >\Delta .$ With lowering the
temperature, this contribution decreases. The contribution caused by Andreev
reflections (the term $g_{A}$ in Eq.(19)) at large $h$ and not small $r_{L}$
is very small as the condensate amplitude is small. Indeed, in this case one
can linearize the Usadel equation (20) and obtain for $f^{R}$

\begin{equation}
f^{R}(x)=f_{S}^{R}/(r_{L}\sqrt{-2ih})\exp (\sqrt{-2ih}(1-x/L))  \label{K22}
\end{equation}

It follows from this equation that at $h_{ex}/E_{Th}=h>>(2r_{L})^{-2}$ the
amplitude of $f^{R}(L)$ is small (we consider low energies $\epsilon
<<\Delta ).$ Therefore the ''Andreev conductance'' $g_{A}$ is also small.
The suppression of the Andreev reflections in the model considered follows
directly from the Usadel equation and the widely used boundary conditions of
Ref. \cite{kuprianov}. It is worth mentioning that the Andreev reflections
are responsible for a subgap conductance in N/I/S junctions, which was
observed in Ref.\cite{kastalsky}. The suppression of the Andreev conductance
by the exchange field is analogous to the suppression of the subgap
conductance by an external magnetic field\cite{vanWees_92,volkov_jetp}. 
%%%%%%%%%%%%%%%%%%%%%%%%%%%%%%%%%%%%%%%%%
%\begin{figure}
%\epsfysize = 8cm
%\vspace{0.2cm}
%\centerline{\epsfbox{./figures/plot1t.eps
% }}
%\vspace{0.2cm}
%\caption{The dependence of the conductance variation $\protect\delta %
%S(T)=S(T)-1$, for different values of the exchange field $h=h_{ex}/E_{Th}$.
%Here $\Delta/E_{Th}=10$, $r_0=1.10^{-3}$ and $r_L=0.1$.}
%\label{fig_plot1}
%\end{figure}
%%%%%%%%%%%%%%%%%%%%%%%%%%%%%%%%%%%%%%%%%%
%%%%%%%%%%%%%%%%%%%%%%%%%%%%%%%%%%%%%%%%%
%
%\begin{figure}[tbp]
%\epsfysize = 8cm
%\vspace{0.2cm}
%\centerline{\epsfbox{./figures/plot2t.eps
%$ }}
% \vspace{0.2cm}
%\caption{The dependence of the conductance variation $\protect\delta %
%S(T)=S(T)-1$, for different values of the exchange field $h=h_{ex}/E_{Th}$.
%Here $\Delta/E_{Th}=10$, $r_0=1.10^{-3}$ and $r_L=2$. The solid, dashed and
%dotted lines correspond to $h=50$, $h=100$ and $h=500$ respectively.}
%\label{fig_plot2}
%\end{figure}
%%%%%%%%%%%%%%%%%%%%%%%%%%%%%%%%%%%%%%%%%%

\section{Triplet component and long-range proximity effect}

In this section, we consider again a ferromagnetic wire with a metallic
contact with a superconductor. In the previous sections we have shown that
the proximity effect in the presence of a strong exchange field $h_{ex}$
could be neglected, since the superconducting condensate penetrates into the
F wire only over a short distance ($\sim \sqrt{\hbar D/h_{ex}}$ in the dirty
limit). At the same time, the interface resistance\ $R_{bs}$ increases when
the normal reservoir becomes superconducting. In short, the total
conductance decreases with decreasing temperature below the superconducting
critical temperature $T_{c}$. 

However, in recent experiments on S/F structures a considerable increase of
the conductance below $T_{C}$ was observed \cite{petrashov,pannetier}. The
measurements also demonstrate that the entire change of the conductance is
due to an increase of the conductivity of the ferromagnetic wire as if the
superconducting condensate penetrated into the ferromagnet. A question arises: how can the condensate function penetrate into the F wire over large
distances? We understand that the Cooper pairs must be destroyed by the
strong exchange field because the electrons of the pairs can no longer have
opposite spins 

To answer this question we notice that such a simple argument about the
destruction of the condensate is based on the assumption that only the
singlet pairing  exists in the sample. This argument cannot be correct if
a triplet component of the superconducting condensate penetrates the
ferromagnet. An arbitrary exchange field cannot destroy the triplet
superconducting component since both the electrons of the Cooper pair are in
the same spin band. In this section, we suggest a mechanism of formation of
the triplet pairing, which is due to a local inhomogeneity of the
magnetization in the vicinity of the S/F interface. We will show that the
penetration length of the triplet component into the ferromagnet is of the
order $\sqrt{\hbar D/\epsilon }$, where the energy $\epsilon $ is of the
order of the temperature $T$ or the Thouless energy $E_{Th}$, and therefore
the increase of the conductance due to the proximity effect may be
comparable with that in a S/N structure.
%%%%%%%%%%%%%%%%
%\begin{figure}[tbp]
%\epsfysize = 7cm
%\centerline{\epsfbox{./figures/geometry_triplet.eps
% }}
%\caption{Schematic view of the structures under consideration.}
%\label{fig_geometry_triplet}
%\end{figure}
%%%%%%%%%%%%%%%%%
We consider the system shown in Fig.\ref{fig_geometry_triplet} and assume
that there is a domain wall in the region $0<x<w$ described by the angle $%
\alpha (x)$ between the magnetization $M$ and the $z$-axis. In this region
the magnetization is given by 
\begin{equation}
{\bf M}=h_{ex}\left( 0,\sin \alpha (x),\cos \alpha (x)\right) \;.
\label{magnetization}
\end{equation}

For simplicity we assume that $\alpha (x)$ varies linearly according to $%
\alpha (x)=Qx$. We consider again the diffusive limit corresponding to a
short mean free path and to the condition $h_{ex}\tau \ll 1$, which allows
us to describe the system using the Usadel equation\cite{usadel}. We assume
that the resistance of the F/S interface is not too small and therefore the
condensate amplitude $|\hat{f}|$ is small. We use the linearized Usadel
equation for the retarded matrix (in spin space) Green function $\hat{f}^{R}$%
, which has the form (the index $R$ is dropped)

\begin{equation}
-iD\partial _{{\bf r}}^{2}\hat{f}+2\epsilon \hat{f}-2\Delta \hat{\sigma}%
_{3}+\left( \hat{f}\hat{V}^{\ast }+\hat{V}\hat{f}\right) =0\;.  \label{a2}
\end{equation}
Here the matrix $\hat{V}$ is defined as $\hat{V}=h_{ex}\left( \hat{\sigma}%
_{3}\cos \alpha (x)+\hat{\sigma}_{2}\sin \alpha (x)\right) $. We also assume
that the diffusion coefficient is the same for the both spin bands. This
equation is supplemented by the boundary condition at the S/F interface,
which after linearization takes the form \cite{kuprianov,zaitsev} 
\begin{equation}
\left. \partial _{x}\hat{f}\right| _{x=0}=\left( \rho /R_{b}\right) F_{S}\;,
\label{bound-cond}
\end{equation}
where $\rho $ is the resistivity of the ferromagnet, $R_{b}$ is the S/F
interface resistance per unit area in the normal state, and $F_{S}=\hat{%
\sigma}_{3}\Delta /\sqrt{\epsilon ^{2}-\Delta ^{2}}$. We have assumed that
there are no spin-flip processes at the S/F interface. 

The solution of Eq. (\ref{a2}) in the region $0<x<w$ can be sought in the
form 
\begin{equation}
\hat{f}=\hat{U}\left( x\right) \hat{f}_{n}\hat{U}\left( x\right) \;,
\label{rotation2}
\end{equation}
where $\hat{U}$ is an unitary transformation given by $\hat{U}\left(
x\right) =\hat{\sigma}_{0}\cos \left( Qx/2\right) +i\hat{\sigma}_{1}\sin
\left( Qx/2\right) $.

Substituting Eq. (\ref{rotation2}) into Eq. (\ref{a2}) we obtain the
following equation for $\hat{f}_{n}$ 
\begin{equation}
-iD\partial _{xx}^{2}\hat{f}_{n}\!\!+\!\!i\left( DQ^{2}/2\right) \left( \hat{%
f}_{n}+\hat{\sigma}_{1}\hat{f}_{n}\hat{\sigma}_{1}\right) \!\!+\!\!DQ\left\{
\partial _{x}\hat{f}_{n},\hat{\sigma}_{1}\right\} +2\epsilon \hat{f}%
_{n}+h_{ex}\left\{ \hat{\sigma}_{3},\hat{f}_{n}\right\} \!=0\;.  \label{a5}
\end{equation}
where $\{...\}$ is the anticommutator. 

In the region $x>w$ the magnetization is homogeneous and $\hat{f}_{n}$
satisfies Eq. (\ref{a5}) with $Q=0$. We see from Eq. (\ref{a5}) that the
singlet and triplet component of the condensate function are mixed by the
rotating exchange field $h_{ex}$. In the region $x>w$ these components
decouple and they should be found by matching the solutions at $x=w$ ( the
function $\hat{f}$ is continuous in the entire F wire). The solution of Eq. (%
\ref{a5}) can be written in the form 
\begin{equation}
\hat{f}_{n}=\hat{\sigma}_{0}A\left( x\right) +\hat{\sigma}_{3}B\left(
x\right) +i\hat{\sigma}_{1}C\left( x\right) \;.  \label{2}
\end{equation}
Here the function $C(x)$ is the amplitude of the triplet component, whereas $%
A(x)$ and $B(x)$ describe the singlet pairing. The structure of Eq. (\ref{a5}%
) allows to seek these amplitudes in the form 
\begin{equation}
A\left( x\right) =\sum_{i=1}^{3}\left( A_{i}\exp \left( -\kappa _{i}x\right)
+\bar{A}_{i}\exp \left( \kappa _{i}x\right) \right)   \label{3}
\end{equation}
The functions $B(x)$ and $C(x)$ can be written in a similar way. The
coefficients $A_{i}$, $B_{i}$ and $C_{i}$ obey the algebraic equations

\begin{eqnarray}
\left( \kappa ^{2}-\kappa _{\epsilon }^{2}-Q^{2}\right) C-2\left( Q\kappa
\right) A &=&0  \nonumber \\
\left( \kappa ^{2}-\kappa _{\epsilon }^{2}\right) B-\kappa _{h}^{2}A &=&0
\label{4} \\
\left( \kappa ^{2}-\kappa _{\epsilon }^{2}-Q^{2}\right) A-\kappa
_{h}^{2}B+2\left( Q\kappa \right) C &=&0\;,  \nonumber
\end{eqnarray}
where $\kappa _{\epsilon }^{2}=-2i\epsilon /D$ and $\kappa
_{h}^{2}=-2ih_{ex}/D$ (indices $i$ were dropped). The eigenvalues $\kappa
_{i}$ are the values at which the determinant of Eqs. (\ref{4}) turns to
zero. In the case of a strong exchange field $h_{ex}$, $\kappa _{h}$ is
large ($\kappa _{h}\gg \kappa _{\epsilon },Q$) and the eigenvalues $\kappa
_{i}$ are given by 
\begin{eqnarray}
\kappa _{1,2} &\approx &(1\pm i)/\xi _{F},\;\;{\rm for}\;\;0<x<L \\
\kappa _{3} &=&\left\{ 
\begin{array}{r@{\quad {\rm for} \quad}l}
\sqrt{\kappa _{\epsilon }^{2}+Q^{2}} & 0<x<w \\ 
\kappa _{\epsilon } & w<x<L
\end{array}
\right. \;.
\end{eqnarray}
The eigenvalues $\kappa _{1,2}$ correspond to a sharp decay of the
condensate in the ferromagnet, while $\kappa _{3}$ is associated with the
slowly varying part. With the boundary condition, Eq. (\ref{bound-cond}), we
see that in a homogeneous case ($Q=0)$ the amplitude of the triplet
component $C(x)$ is zero. If $Q\not=0,$ the function $C(x)$ is coupled to
the singlet components $A(x)$ and $B(x),$ and we consider this case. 

If the width $w$ is small, the triplet component changes only a little in
the region $(0,w)$ and spreads over a large distance of the order $\left|
\kappa _{\epsilon }^{-1}\right| $ in the region $(0,L)$. In the case of a
strong exchange field $h_{ex}$, $\xi _{F}$ is very short ($\xi _{F}\ll w,\xi
_{T}$), the singlet component decays very fast over the length $\xi _{F}$,
and its slowly varying part $B_{3}$ is $B_{3}=2\left( Q\kappa _{3}/\kappa
_{h}^{2}\right) C_{3}\ll C_{3}$. In order to obtain the expression for the
triplet component $C(x),$ we use Eq. (\ref{bound-cond}) at $x=0$ and assume
that the solution vanishes at $x=L.$ Then, we find 
\begin{equation}
C^{R(A)}(x)=\mp i\left\{ QB(0)\sinh \left( \kappa _{\epsilon }(L-x)\right) 
\left[ \kappa _{\epsilon }\cosh \Theta _{\epsilon }\cosh \Theta _{3}+\kappa
_{3}\sinh \Theta _{\epsilon }\sinh \Theta _{3}\right] ^{-1}\!\right\}
^{R(A)}\;,  \label{5}
\end{equation}
where $w<x<L$, $B^{R(A)}(0)\!\!\!=\!\!\left( \rho \xi _{h}/2R_{b}\right)
f_{S}^{R(A)}$ is the amplitude of the singlet component at the S/F
interface, $\Theta _{\epsilon }=\kappa _{\epsilon }L$, $\Theta _{3}=\kappa
_{3}w$, and $\kappa _{\epsilon }^{R(A)}=\sqrt{\mp 2i\epsilon /D}$. 

\begin{figure}[tbp]
\epsfysize = 12cm
\centerline{\epsfbox{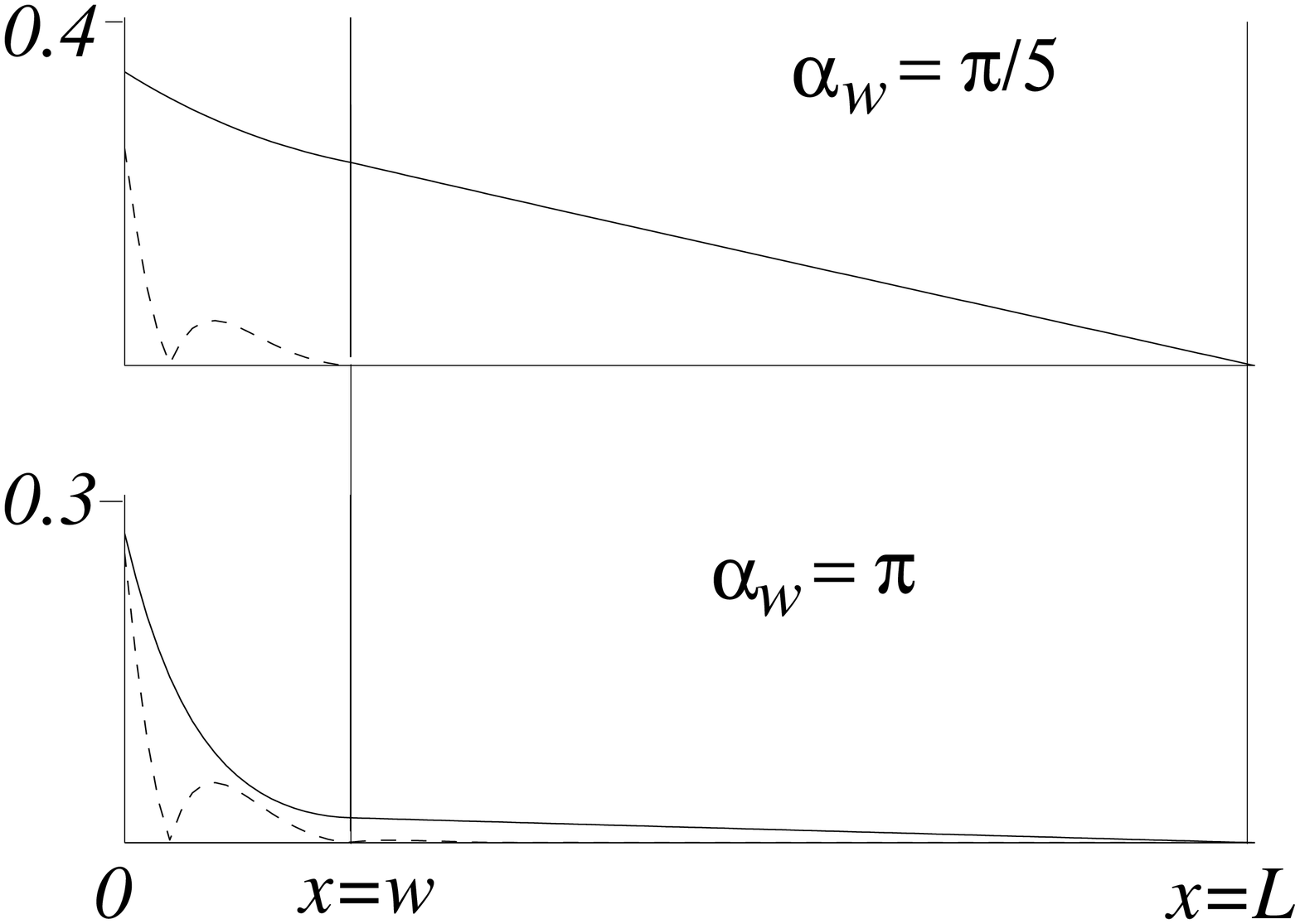}}
\vspace{2cm}
\caption{F. S. Bergeret, A.F. Volkov and K.B.Efetov. ``Transport and triplet superconducting condensate in mesoscopic
ferromagnet-superconductor structures.''}
\label{fig_penetration}
\end{figure}

It is clear from Eq.(\ref{5}) that, at the interface, the triplet component
is of the same order of magnitude as the singlet one. Indeed, for the case $%
w\ll L$ we obtain from Eq. (\ref{5}) $\left| C(0)\right| \sim B(0)/\sinh
\alpha _{w}$, where $\alpha _{w}=Qw$ is the angle characterizing the
rotation of the magnetization. Therefore, provided the angle $\alpha
_{w}\leq 1$ and the S/F interface transparency is not too small, the singlet
and triplet components are not small. They are of the same order in the
vicinity of the S/F interface but, while the singlet component decays fast
over a short distance ($\sim \xi _{F}$), the triplet one varies smoothly
along the ferromagnet turning to zero at the F reservoir (see Fig.\ref
{fig_penetration}). One can see also that the singlet component oscillates,
which is well known \cite{buzdin_kupr1}.

\begin{figure}[tbp]
\epsfysize = 14cm
\centerline{\epsfbox{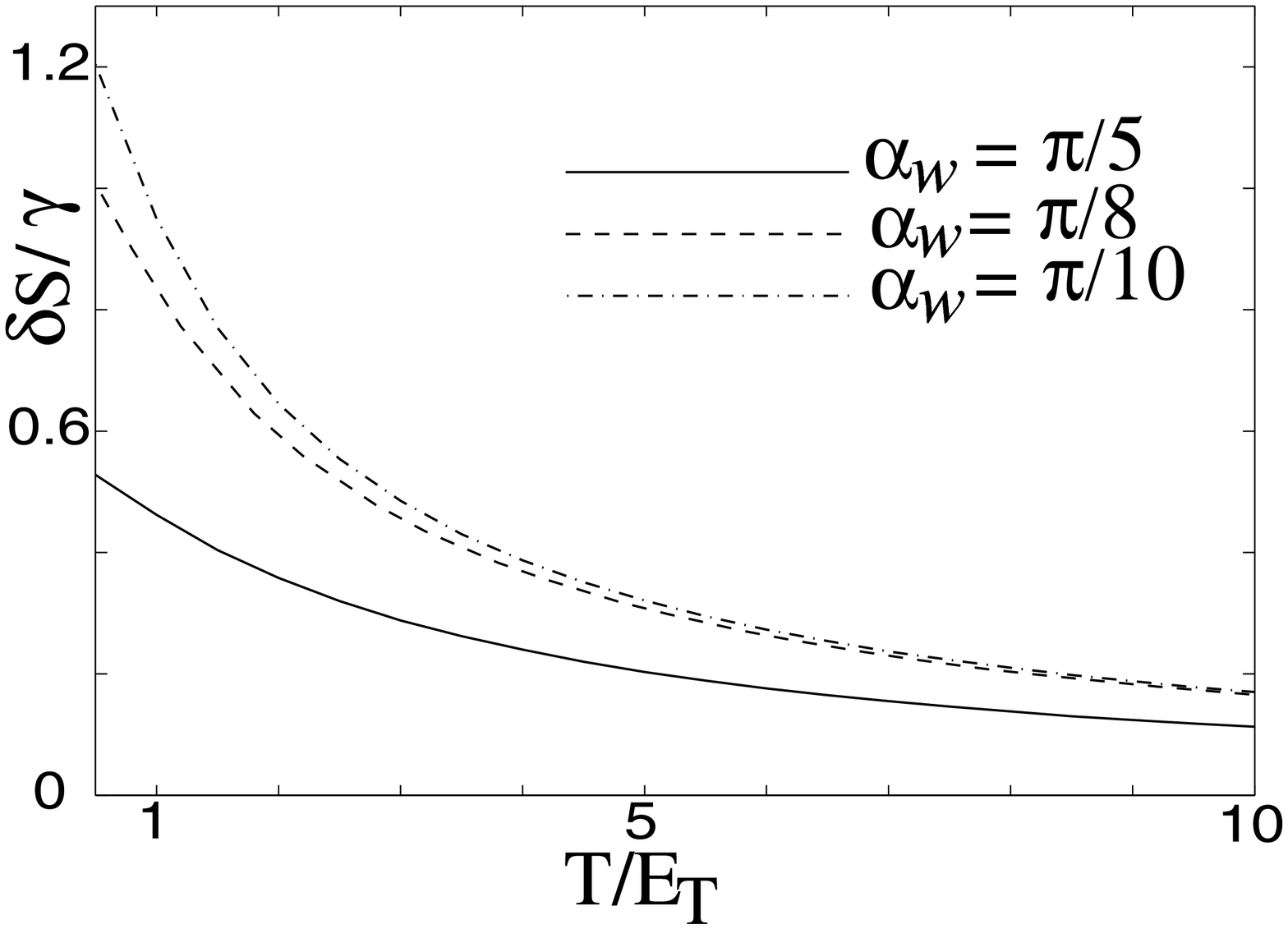}}
\vspace{2cm}
\caption{F. S. Bergeret, A.F. Volkov and K.B.Efetov. ``Transport and triplet superconducting condensate in mesoscopic
ferromagnet-superconductor structures.''}
\label{fig_conductance_q}
\end{figure}

%%%%%%%%%%%%%%%%
%\begin{figure}[tbp]
%\epsfysize = 7cm
%\centerline{\epsfbox{./figures/penetration.eps}}
%\caption{Spatial dependence of the singlet (dashed line) and the triplet
%(solid line) components of $|\hat{f}|$ in the F wire for different values of 
%$\protect\alpha _{w}$. Here $w=L/5$, $\protect\epsilon =E_{T}$ and $%
%h/E_{T}=400$. $E_{T}=D/L^{2}$ is the Thouless energy. }
%\label{fig_penetration}
%\end{figure}
%%%%%%%%%%%%%%%%%
The penetration of the triplet component into the ferromagnet is similar to
the penetration of the superconducting condensate into a normal metal. The
presence of the condensate function (triplet component) in the ferromagnet
can lead to long-range effects and therefore to a significant change of the
conductance of a ferromagnetic wire in a S/F structure (see inset in Fig.\ref
{fig_geometry_triplet}) when the temperature decreases below $T_{c}$. The
normalized conductance variation $\delta S=\left( G_{s}-G_{n}\right) /G_{n}$
is given by the expression \cite{VZK,volkov_taka}: 
\begin{equation}
\delta S=-\frac{1}{32T}{\rm Tr}\int {\rm d}\epsilon F_{V}^{\prime
}\left\langle \left[ \hat{f}^{R}(x)-\hat{f}^{A}(x)\right] ^{2}\right\rangle
\;.  \label{6}
\end{equation}
Here $G_{n}$ is the conductance in the normal state, $<..>$ denotes the
average over the length of the ferromagnetic wire between the F reservoirs,
and $F_{V}^{\prime }$ is given by the expression 
\begin{equation}
F_{V}^{\prime }=1/2\left[ \cosh ^{-2}((\epsilon +eV)/2T)+\cosh
^{-2}((\epsilon -eV)/2T)\right] \;.
\end{equation}
Substituting Eqs. (\ref{2}, \ref{5}) into Eq. (\ref{6}) one can determine
the temperature dependence $\delta S\left( T\right) $ shown in Fig.\ref
{fig_conductance_q}.
 %%%%%%%%%%%%%%%%
%\begin{figure}[tbp]
%\epsfysize = 8.5cm
%\centerline{\epsfbox{./figures/conductance_q.eps}}
%\caption{The $\protect\delta S(T)$ dependence. Here $\protect\gamma =\protect%
%\rho \protect\xi _{h}/R_{b}$. $\Delta /E_{T}\gg 1$ and $w/L=0.05$.}
%\label{fig_conductance_q}
%\end{figure}
%%%%%%%%%%%%%%%%%
We see that $\delta S$ increases with decreasing temperature and saturates
at $T=0$. In order to explain the reentrant behavior of $\delta S(T)$
observed in Refs. \cite{petrashov,pannetier}, one should take into account
other mechanisms as those analyzed in Refs. \cite
{falko_volkov1,golubov,BVE_prl2} and in section 2. However, this question is
beyond the scope of the present analysis.

It is also interesting to note that a triplet component of the condensate
function with the same symmetry (odd in frequency $\omega $ and even in
momentum $p$) has been suggested long ago by Berezinskii \cite{berezin} as a
possible phase in superfluid $^{3}He$ (this, so called ``odd
superconductivity'', was discussed in a subsequent paper \cite{balat}).
Being symmetric in space, this component is not affected by potential
impurities, in contrast to the case analyzed in Ref. \cite{larkin_triplet},
where the triplet component of the condensate was odd in space. While this
hypothetical condensate function is not realized in $^{3}He$ (in $^{3}He$ it
is odd in $p$ but not in frequency), this odd (in $\omega $) triplet
component does exist in the system considered here, although under special
conditions described above.

As it was mentioned in the introduction, experimental data are still
controversial. It has been established in an experiment \cite{chandrasekhar}
that the conductance of the ferromagnet does not change below $T_{c}$ and
all changes in $\delta S$ are due to changes of the S/F interface resistance 
$R_{b}$. However, in other experiments $R_{b}$ was negligibly small \cite
{petrashov}. The mechanism suggested in our work may explain the long-range
effects observed in the experiments \cite{petrashov,pannetier}. At the same
time, the result of the experiment \cite{chandrasekhar} is not necessarily
at odds with our findings. The inhomogeneity of the magnetic moment at the
interface, which is the crucial ingredient of our theory, is not a
phenomenon under control in these experiments. One can easily imagine that
such inhomogeneity existed in the structures studied in Refs. \cite
{petrashov,pannetier} but was absent in those of Ref. \cite{chandrasekhar}.
The magnetic inhomogeneity near the interface may have different origins.
Anyway, a more careful study of the possibility of a rotating magnetic
moment should be performed to clarify this question.

\section{Conclusion}

We have analyzed the conductance variation $\delta G_{s}\equiv G_{s}-G_{n}$
of the F/S mesoscopic structure in the dirty limit when the condition $%
h_{ex}<\hbar /\tau $ is satisfied. First, neglecting the proximity effect we
have shown that below $T_{c}$ the conductance variation $\delta G_{s}$ is
negative and its magnitude increases with increasing the difference between
the conductances for spin up and down. The change of the conductance is
related to changes in the boundary conditions at the F/S interface. Below $%
T_{c}$ one of the boundary conditions requires the spin current to be zero.
The formulas for the conductance remain valid even if inelastic scattering
(not spin flip) processes are taken into account. 

We also studied how the proximity effect affects the conductance for an
arbitrary transparency of the F/S interface. The account for the condensate
penetration (the proximity effect) leads to an increase of the conductance
near $T_{c}$. This is possible, however, only if the exchange energy $h_{ex}$
is not too large: the parameter $hr_{L}^{2}$ should not be large, where $h$
is the exchange energy in units of the Thouless energy $\epsilon _{Th}=\hbar
D/L^{2}$ and $r_{L}$ is the ratio of the F/S interface resistance in the
normal state to the resistance of the F wire. If the parameter $hr_{L}^{2}$
is large, the conductance $G_{s}$ decreases with decreasing temperature and
becomes very small at low $T.$ This behavior is related to a strong
suppression of the Andreev reflection processes that determine the
conductance at low $T$ by the exchange field. 

It is important to note that we neglected the inverse proximity effect, that
is, suppression of superconductivity in S. In the case of large $h$ one has
to take into account this effect and calculate a deviation of the Green's
functions in the superconductor from their bulk values. It turns out that in
the case of a large exchange energy $h,$ the energy gap in the
superconductor may be suppressed and the DOS $\nu _{S}$ is not zero even at 
%TCIMACRO{\TEXTsymbol{\vert}}%
%BeginExpansion
\mbox{$\vert$}%
%EndExpansion
$\epsilon $%
%TCIMACRO{\TEXTsymbol{\vert}}%
%BeginExpansion
\mbox{$\vert$}%
%EndExpansion
$<\Delta $. In this case the quasiparticle conductance is not zero and the
charge transfer is possible through the F/S interface at low energies $%
\epsilon .$ The conversion of the quasiparticle current into the current of
Cooper pairs is realized in the superconductor over the coherence length $%
\xi _{S}$ (at low energies).

At last, we have calculated the conductance of a F/S structure in which the
magnetization is non-homogeneous near the F/S interface. It was shown that,
in this case, a triplet component arises alongside with the singlet one.
This triplet component in another systems (He$^{3}$, high $T_{c}$
superconductors etc) was studied in several papers \cite{balat,berezin}; it
is odd in Matsubara frequencies (the so called odd superconductivity) and
even in momentum space (in the main approximation in the parameter ($\hbar
/h_{ex}\tau $)). The triplet component penetrates into the F wire over a
large distance of the order $\sqrt{\hbar D/T}$ and leads to a positive
conductance variation. The singlet component penetrates over much shorter
length of the order $\sqrt{\hbar D/h_{ex}}$ and leads to a negligible
contribution to the conductance. It would be interesting to realize
experimentally the situation that we analyzed in the last section and to
observe the triplet component not in an exotic system but in an ordinary F/S
structure.

We would like to thank SFB 491 for financial support.

\bigskip

\begin{appendix}

\section{}

Here we show how the boundary conditions for the distribution functions $%
N_{\alpha }=n_{\alpha }-p_{\alpha }$ can be derived from matching conditions
for the quasiclassical Green functions, where $n_{\alpha }$ is the
distribution function of electrons and $p_{\alpha }=1-n_{\alpha }(-\epsilon
)$
is the distribution functions of holes (having in mind a superconductor, it
is better to speak about electron- and hole-like excitations). We use the
boundary conditions for the quasiclassical matrix Green functions $g$ the
matrix elements of which consist of the retarded (advanced) Green's
functions $\ghat^{R(A)}$ and the Keldysh Green's function $\ghat^{K}.$ The
last
matrix is expressed in terms of the matrix distribution function $\fhat:$ $%
\ghat^{K}=\ghat^{R}\fhat-\fhat\ghat^{A}$, where the matrix $\fhat$ can be
represented in the form: $%
\fhat=f_{1}\hat1+\fhat_{3}\hat{\sigma} _{3}.$ The first function $f_{1}$
determines the order
parameter in the superconductor and is equal $f_{1\alpha }=1-(n_{\alpha
}+p_{\bar{\alpha}}).$ The second function $f_{3}$ determines the electrical or
spin current and equals $f_{3\alpha }=-(n_{\alpha }-p_{\bar{\alpha} }).$ The
matching condition at the F(N)/S interface for the 4$\times $4 matrix
Green function has the form \cite{kuprianov}

\begin{equation}
L\gcheck\partial _{x}\gcheck=(1/2r_{L})[\gcheck,\gcheck_{S}]\;.  \label{A1}
\end{equation}

where $r_{L}=R_{b}/\rho L$ is the F/S interface resistance (in the normal
state) normalized to the resistance of the F wire $\rho L.$ The physical
meaning of Eq.(A1) is simple. If we take the Keldysh (the element (12)) of
Eq.(A1), multiply it by $\hat{\sigma}_{3}$ and calculate the trace, we
obtain the current at a given energy $\epsilon $. If we integrate over
energies,  we obtain on the left side the usual expression for the current

\begin{equation}
I_{Q}=(1/4)e\nu _{n}D{\rm Tr}\hat{\sigma}_3\int {\rm d}\epsilon \lbrack
\ghat^{R}\partial
_{x}\ghat^{K}+\ghat^{K}\partial _{x}\ghat^{A}]  \label{A2}
\end{equation}

where $\nu _{n}$ is the DOS in the normal state. On the right hand side we obtain
the current at a given energy which is well known in the tunnel Hamiltonian
method. We take the Keldysh component of Eq. (A1), multiply it by $\hat{%
\sigma}_{3}$ and $\hat{1}$ and calculate the trace. We obtain thus the
equations

\begin{equation}
M_{3}\partial
_{x}f_{3}=(1/r_L)\left[\left(\nu\nu_s+g_a\right)f_3-\left(g_-f_{eq}+g_+f_1\right)\right]
\label{A3}
\end{equation}

\begin{equation}
M_{1}\partial
_{x}f_{1}=(1/r_L)\left[\left(\nu\nu_s+g_1\right)\left(f_1-f_{eq}\right)-g_-f_3\right]
\label{A4}
\end{equation}

where \ \ $\nu =%
%TCIMACRO{\func{Re}}%
%BeginExpansion
\mathop{\rm Re}%
%EndExpansion
g^{R}$, \ $\nu _{S}=%
%TCIMACRO{\func{Re}}%
%BeginExpansion
\mathop{\rm Re}%
%EndExpansion
G_{S}^{R}$ is the DOS in the ferromagnet and superconductor, respectively
and

$M_{3}=(1/2)[1-g^Rg^A-f^Rf^A]$

$M_{1}=(1/2)[1-g^Rg^A+f^Rf^A]$

$g_{1}=-(1/4)(f^R-f^A)(F^R_s-F^A_s)$

$g_{\pm }=+(1/4)\{(\fhat^R\mp\fhat^A)(\hat{F}^R_s\pm\hat{F}^A_s)\}_3$

$g_{A}=-(1/4)(f^R+f^A)(F^R_s+F^A_s)$

The symbol \{...\}$_{3}$ stands for $\{...\}_{3}=(1/2)Tr\hat{\sigma}%
_{3}\{...\}$. In the case under consideration $f^{R(A)}\propto i$
$\hat{\sigma%
}_{2}$ and $F_{S}^{R(A)}\propto i$ $\hat{\sigma}_{2},$so that $g_{\pm }=0$.
We are interested in the states with subgap energies (low temperatures) $%
%TCIMACRO{\UNICODE[m]{0xa6}}%
%BeginExpansion
{\vert}%
%EndExpansion
\epsilon
%TCIMACRO{\UNICODE[m]{0xa6}}%
%BeginExpansion
{\vert}%
%EndExpansion
<\Delta .$ For these states one has $F_{S}^{R}=F_{S}^{A}=\Delta /i\sqrt{%
\Delta ^{2}-\epsilon ^{2}}$ and $G_{S}^{R}=G_{S}^{A}=\epsilon /i\sqrt{\Delta
^{2}-\epsilon ^{2}}.$ Therefore we obtain that $g_{1}=0,$ and \ $\nu
_{S}=0,$
but $g_{A}\neq 0.$ One can show that at any non-zero $r_{L}$ the functions
$%
M_{3}$ and $M_{1}$ are not zero and we find from Eqs.(A3-A4)

\begin{equation}
\partial _{x}f_{1}=0\Longrightarrow\left\{ \begin{array}{l}
\partial_x(n_\uparrow+p_\downarrow)=0\\
\partial_x(n_\downarrow+p_\uparrow)=0 \end{array}\right.\label{A5}
\end{equation}

From Eq.(A5) we find the boundary condition

\begin{equation}
\partial _{x}N_{\uparrow }-\partial _{x}N_{\downarrow }=0  \label{A6}
\end{equation}

This condition means the absence of the spin current. It may be easily
generalized to the case of different diffusion coefficients

\begin{equation}
D_{\uparrow }\partial _{x}N_{\uparrow }-D_{\downarrow }\partial
_{x}N_{\downarrow }=0  \label{A7}
\end{equation}

At small $r_{L}$ (a good F/S contact) the condition (A3) yields
\begin{equation}
f_{3}=0\Longrightarrow \left\{ \begin{array}{l}
n_\uparrow-p_\downarrow=0\\ n_\downarrow-p_\uparrow=0 \end{array}\right.
\label{A8}
\end{equation}

We took into account that in the sub-gap region the ''Andreev'' conductance
$%
g_{A}$ is not zero. From Eq.(A8) we get

\begin{equation}
N_{\uparrow }=-N_{\downarrow }  \label{A9}
\end{equation}

In the normal state the distribution function is continuous across a good
F/S interface. Hence we get $n_{\alpha }=p_{\bar{\alpha}}\Longrightarrow
N_{\alpha }=0.$ We use these boundary conditions in section 2.
\end{appendix}

\clearpage

\underline{Figure Captions}

Fig.1: S/F system.

\bigskip

Fig.2: The dependence of the conductance variation $\protect\delta %
S(T)=S(T)-1$, for different values of the exchange field $h=h_{ex}/E_{Th}$.
Here $\Delta/E_{Th}=10$, $r_0=1.10^{-3}$ and $r_L=0.1$.

\bigskip

Fig.3: The dependence of the conductance variation $\protect\delta %
S(T)=S(T)-1$, for different values of the exchange field $h=h_{ex}/E_{Th}$.
Here $\Delta/E_{Th}=10$, $r_0=1.10^{-3}$ and $r_L=2$. The solid, dashed and
dotted lines correspond to $h=50$, $h=100$ and $h=500$ respectively.

\bigskip 

Fig.4: Schematic view of the structures under consideration.

\bigskip 

Fig.5: Spatial dependence of the singlet (dashed line) and the triplet
(solid line) components of $|\hat{f}|$ in the F wire for different values of 
$\protect\alpha _{w}$. Here $w=L/5$, $\protect\epsilon =E_{T}$ and $%
h/E_{T}=400$. $E_{T}=D/L^{2}$ is the Thouless energy.

\bigskip 

Fig.6:  The $\protect\delta S(T)$ dependence. Here $\protect\gamma =\protect%
\rho \protect\xi _{h}/R_{b}$. $\Delta /E_{T}\gg 1$ and $w/L=0.05$.

\clearpage

\clearpage

%%%%%%%%%%%%%%%%%
%%%%%%%%%%%%%%%%%

\clearpage
%%%%%%%%%%%%%%%%%%%%%%%%%%%%%%%%%%%%%%%%%
%%%%%%%%%%%%%%%%%%%%%%%%%%%%%%%%%%%%%%%%%%

\clearpage

%%%%%%%%%%%%%%%%%%%%%%%%%%%%%%%%%%%%%%%%%

%%%%%%%%%%%%%%%%%%%%%%%%%%%%%%%%%%%%%%%%%%%%%%%%%%%%

\clearpage

%%%%%%%%%%%%%%%%
%%%%%%%%%%%%%%%%%

\clearpage

%%%%%%%%%%%%%%%%
%%%%%%%%%%%%%%%%%

\clearpage

 %%%%%%%%%%%%%%%%
%%%%%%%%%%%%%%%%%

%\bibliographystyle{prsty.bst}
%\bibliography{list}

\end{document}